\documentstyle[pre,aps,epsf,amssymb,mathptm]{revtex}
\begin{document}
\preprint{TIFR/TH/01--18}
\draft

\title{Probability distribution of the sizes of largest erased-loops in
loop-erased random walks}

\author{Himanshu Agrawal\cite{HA} and Deepak Dhar\cite{dd}}
\address{Department of Theoretical Physics, Tata Institute of Fundamental
	 Research, Homi Bhabha Road, Colaba, Mumbai -- 400 005, India}
\date{July 12, 2001}

%
% REMOVE THE FOLLOWING LINE IN THE FINAL VERSION
%
\twocolumn[\hsize\textwidth\columnwidth\hsize\csname@twocolumnfalse\endcsname
\maketitle
\begin{abstract}

We have studied the probability distribution of the perimeter and the area of
the $k$th largest erased-loop in loop-erased random walks in two-dimensions for
$k$ = $1$ to $3$. For a random walk of $N$ steps, for large $N$, the average
value of the $k$th largest perimeter and area scales as $N^{5/8}$ and $N$
respectively. The behavior of the scaled distribution functions is determined
for very large and very small arguments. We have used exact enumeration for $N
\leq 20$ to determine the probability that no loop of size greater than $\ell$
is erased. We show that correlations between loops have to be taken into account
to describe the average size of the $k$th largest erased-loops. We propose a
one-dimensional Levy walk model which takes care of these correlations. The
simulations of this simpler model compare very well with the simulations of the
original problem.

\end{abstract}
\pacs{05.40.Fb, 05.65.+b, 02.50.-r, 02.70.Uu}

%05.40.Fb Random walks and Levy flights
%05.65.+b Self-organized systems
%02.50.-r Probability theory, stochastic processes, and statistics
%02.70.Uu Applications of Monte Carlo methods
%
% REMOVE THE FOLLOWING LINE IN THE FINAL VERSION
%
\vskip3ex]

\narrowtext

\section{INTRODUCTION}
\label{Sec:Intro}

The classical theory of statistics of extremals deals with the distribution of
the largest of many {\em independent identically distributed\/} random variables
\cite{lead83,gned43}. After some rescaling, this distribution tends to one of
the three universal distribution functions, the so-called Gumbel distributions,
independent of the details of the starting distribution of random variables
\cite{gumnote}. The independence of  random variables is, however, not a good
approximation in many physical problems. The statistics of extremes of many
random variables is relevant in many different physical contexts. In many of
these it is important to take account of correlations, for example, in the study
of earthquakes \cite{sorn96}, weather records \cite{schm99}, slow relaxation in
glassy systems \cite{ledo00}, and persistence in random walks \cite{rieg98}. In
some special cases extremal statistics of strongly correlated variables can be
determined exactly \cite{krap00}. In general, however, the study of extremal
distributions of correlated and strongly correlated random variables poses a
rather non-trivial problem even in the simplest cases. 

This paper deals with the extremal statistics of variables with long-range power
law correlations in the loop-erased random walks (LERW's) in two dimensions. Our
interest in the LERW problem comes from the fact that it provides one of the
simplest examples of self-organized critical systems. In the LERW problem, the
length of the walk is first increased by one at each step, and then decreases by
a random amount due to possible loop erasures. The size of erased loops has a
power-law tail \cite{dd97}. This is, thus, similar to the sandpile model where
one grain is added at each time step but the number of grains leaving the pile
has a power-law tail. Clearly, there are correlations in the sizes of erased
loops at different times. These correlations are more pronounced for larger
loops. Erasure of a large loop leads to significant decrease in the length of
the erased walk, and hence a significant decrease in the probability of erasure
of another large loop within a short time. We propose that the expected ratios
of sizes of $k$th largest loop with the largest loop is a good variable to
quantify these strong correlations, and propose a one-dimensional Levy walk
model which is then tested by simulations.

The LERW problem was introduced by Lawler \cite{law80} as a more tractable
variant of the self-avoiding walk problem. This problem is related to many
well-studied problems in statistical physics: the classical graph-theoretical
problem of spanning trees, the $q$-state Potts model in the limit $q \to 0$
\cite{maj92}, and the Laplacian self-avoiding walk problem \cite{law87}.
Connection to the spanning trees also relates this problem to the abelian
sandpile model of self-organized criticality \cite{dd99}. Recently simulation of
LERW has been used as a computationally efficient way to determine the dynamical
exponent of the abelian sandpile model in three-dimensions \cite{hadd01}. The
upper critical dimension of LERWs is known to be 4 \cite{lawbook}. In two
dimensions, the fractal dimension of LERWs is known to be $5/4$
\cite{maj92,gut90,ken00}, and the exponent characterizing the probability
distribution of the area of erased-loops is known to be superuniversal
\cite{hadd01}. Several other results on LERWs can be found in
\cite{dd97,law98,dup92,ktit00} and a good review of earlier results on the LERW
problem can be found in \cite{lawbook}.

In this paper, we show that the asymptotic behavior of the probability
distribution function $\text{Prob}(\ell^{(k)} | N)$ that the $k$th largest
erased-loop perimeter in the first $N$ steps has value $\ell^{(k)}$ is described
by a $k$-dependent scaling function with argument $\ell^{(k)}/N^{z/2}$. We
determine the behavior of the scaling function for the largest loop for very
large and very small values of its argument. A similar behavior is found for the
loops ranked by the enclosed area, rather than by their perimeter. The
probability that there is no erased loop of length greater than a fixed value
$r$ varies exponentially with $N$ for large $N$. Enumerating all walks
satisfying this property (for a fixed $r$) is a generalization of the
self-avoiding walk problem. We have used exact enumeration techniques to
determine the behavior of this probability for $r = 0$, $2$, and $4$ by
enumerating all random walks with $N \leq 20$. We have proposed a simple Levy
walk model which captures correlations in the LERW and agrees well with its
extremal statistics as determined from large-scale Monte Carlo simulations.

The plan of this paper is as follows: In Sec.~\ref{Sec:Defs}, after defining the
LERW model precisely, we recall relevant points from scaling theory for
distribution of sizes of erased-loops. These are used to get the scaling form
for the probability distribution of the perimeter and the area of largest
erased-loop in a walk of $N$ steps. In Sec.~\ref{Sec:Mu}, we outline our results
about the connectivity constants $\mu_{2}$ and $\mu_{4}$ and determine their
numerical value using series expansions. The simulation technique and results
obtained thereof are described in Sec.\ \ref{Sec:Sim}. In Sec.\ \ref{Sec:Corr},
we describe the Levy walk model and compare the results of numerical simulations
of this model with that of the LERW. Finally, some concluding remarks follow in
Sec.\ \ref{Sec:Conc}.

\section{Scaling Theory of loop-size Distributions }\label{Sec:Defs}

\begin{figure}
\centerline{\epsfxsize0.7\hsize\epsfbox{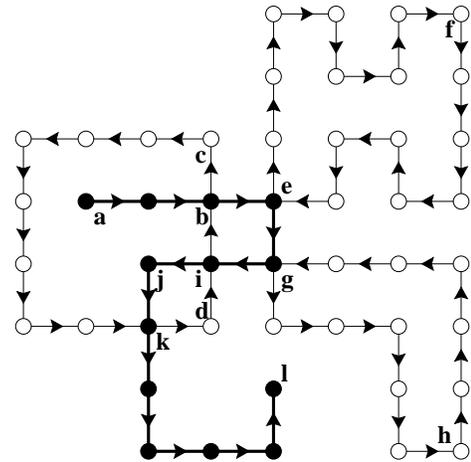}}
\vskip1ex

\caption{An illustrative example of the loop-erasure procedure and some aspects
related to perimeter and enclosed area of erased-loops in loop-erased random
walks: The random walk {\bf a-b-c-d-i-b-e-f-e-g-h-g-i-j-k-l} of 52 steps
starts at {\bf a}, and ends at {\bf l}. The erased-loops are shown by thin lines
and the loop-erased walk {\bf a-b-i-j-k-l} having 12 steps is shown by thick
lines with sites on it marked by solid circles. Note that at the points {\bf i}
and {\bf k}, while the random walk path intersects itself, the LERW encounters
no intersection as the loop {\bf b-c-k-d-i-b} has already been erased.}

\label{F:LERW}
\end{figure}

A loop-erased random walk is defined recursively as follows: For a one step
random walk, the corresponding loop-erased random walk is the same as the random
walk. To form the LERW ${\cal L}'$ corresponding to a given random walk of
$(N+1)$ steps, we first form the LERW ${\cal L}$ corresponding to the first $N$
steps of the random walk. Let us say this LERW ${\cal L}$ has $n$ steps. We now
add the $(N+1)$th step of the random walk to ${\cal L}$. If no loop is formed,
the resulting $n+1$ stepped walk is ${\cal L}'$. If this results in forming a
loop of perimeter $\ell$, this loop is erased, and the resulting $n+1-\ell$
stepped walk is ${\cal L}'$. A simple example is depicted in Fig.~\ref{F:LERW}.

Let ${\cal L}$ be a LERW of $n$ steps obtained from a random walk of $N$ steps.
For a fixed $N$, $n$ is a random variable. The critical exponent $z$  of the
LERW is defined by the relation that
\begin{equation} \label{E:nNz}
  \langle n\rangle  \sim N^{z/2}
\end{equation}
for large $N$, where the angular brackets denote ensemble averaging over all
random walks of $N$ steps. Since the root-mean-square end to end distance $R$
for LERWs is the same as that for random walks, we have $R \sim N^{1/2}$, and
$\langle n \rangle \sim R^{z}$. Thus, $z$ is the fractal dimension of the LERW.

Let $F(\ell | N)$ be the cumulative probability that a loop of perimeter {\em
greater than\/} $\ell$ will be erased {\em at\/} the $N$th step of the
loop-erased walk. We define
\begin{equation}
  F(\ell) = \lim_{N \to \infty} F(\ell | N).
\end{equation}
It was shown in \cite{hadd01} that large $N \gg \ell \gg 1$, $F(\ell | N)$
satisfies the scaling form
\begin{equation}\label{E:FlN}
  F(\ell | N) \sim \ell^{-2/z} f\bigl(\ell/N^{z/2}\bigr).
\end{equation}
The scaling function $f(x)$ tends to a non-zero constant as $x$ tends to zero,
and decreases to zero exponentially fast for $x \gg 1$. Note that the exponents
appearing in this scaling form depend only on the fractal dimension $z$.

Let $\Phi(\ell^{(1)} \le \ell | N)$ be the cumulative probability that the
perimeter $\ell^{(1)}$ of the largest erased-loop in an $N$-step walk will be
{\em less than or equal\/} to $\ell$. We shall study the behavior of this
function for large $N$. The probability that the largest erased-loop at the
$k$th step of an $N$-step walk has perimeter {\em less than or equal\/} to
$\ell$ is given by $1-F(\ell | k)$. A simple approximate formula for
$\Phi(\ell^{(1)} \le \ell | N)$ is obtained by neglecting correlations among
sizes of erased-loops, and treating the generation of loops at different time
steps as independent events. In the following, we will denote by
$\Phi_{\text{uc}}$  the value of $\Phi(\ell^{(1)} \le \ell | N)$ in this
uncorrelated approximation. This gives
\begin{equation}\label{E:PhiF}
  \Phi(\ell^{(1)} \le \ell | N)
    \simeq \Phi_{\text{uc}}(\ell^{(1)} \le \ell | N)
    = \prod_{k=1}^{N} [1-F(\ell | k)].
\end{equation}

Let $x = \ell/N^{z/2}$, $x^{(1)} = \ell^{(1)}/N^{z/2}$ and $y=k/N$ be new
scaling variables. In terms of these new variables, substitution of $F(\ell |
k)$ from Eq.~(\ref{E:FlN}) in Eq.~(\ref{E:PhiF}), gives
\begin{equation}\label{E:PhiXY}
  \Phi_{\text{uc}}(x^{(1)} \le x | N) =
    \prod_{y}
    \left[1-\frac{1}{N}x^{-2/z}f\left(\frac{x}{y^{z/2}}\right)\right].
\end{equation}

For fixed $x$ and large, we can evaluate this expression by taking logs,
expanding in powers of $(1/N)$, and keeping only the lowest order terms in
$(1/N)$. With this we get
\begin{equation}\label{E:logPhi}
  \ln\Phi_{\text{uc}}(x^{(1)} \le x | N) = -x^{-2/z} \tilde{f}(x),
\end{equation}
where $\tilde{f}(x) = \int_{0}^{1} f\left(x/y^{z/2}\right) dy$. It is easy to
see that $\tilde{f}(x)$ has the same qualitative behavior as $f(x)$.
In terms of $F(\ell | k)$, this equation can be written as
\begin{equation}\label{E:PhiExp}
  \Phi_{\text{uc}}(\ell^{(1)} \le \ell | N) \simeq
    \exp\left[-N\tilde{F}(\ell | N)\right],
\end{equation}
where
\begin{equation}
  \tilde{F}(\ell | N) = \frac{1}{N} \sum_{k = 0}^{N}F(\ell | k).
\end{equation}
For small $x$, $\ln\Phi_{\text{uc}}(x^{(1)} \le x | N)$ should vary as
$-x^{-2/z}$. For large $x$, $\tilde{f}(x)$ is small, and $1 -
\Phi_{\text{uc}}(x^{(1)} \le x | N)$ should vary as $x^{-2/z} \tilde{f}(x)$. 

Eq.~(\ref{E:PhiExp}) is a good approximation to Eq.~(\ref{E:PhiF}) so long as
the higher order terms in $(1/N)$ can be neglected. It is easily seen that the
neglected term is of order $N\tilde{F}^{2}(\ell | N)$, and hence the
approximation is valid so long as $\ell \gg N^{z/4}$. It will be seen from
simulation results (see Sec.~\ref{Sec:Sim}) that our assumption about
correlations being small is not too bad and that Eq.~(\ref{E:PhiF}), and
consequently also Eqs.~(\ref{E:logPhi}) and (\ref{E:PhiExp}), are reasonable
approximations to the largest loop size distribution for all $\ell$. The
deviation of the correct value from Eq.~(\ref{E:PhiF}) is largest if $\ell$ is
very small, say equal to $0$, $2$, $4$, \ldots. It is important to understand
the behavior of $\Phi(\ell^{(1)} \le \ell | N) $ in this case. This we do in the
next section.

Let $m$ be the expected number of loops of perimeter {\em greater than or
equal\/} to $\ell$ generated from a random walk of $N$ steps. If there are no
correlations between different loops, for $m \ll N$, the number of such loops
generated in particular realization is a random variable, distributed according
to the Poisson distribution: The probability that exactly $k$ such loops are
generated is  $\text{e}^{-m}m^{k}/k!$. This implies that the probability
that less than $k$ loops of size greater than $\ell$ are generated can be
expressed in terms of the probability that {\em no\/} loop of size greater than
$\ell$ is generated. Simple algebra gives
\begin{equation}
\label{E:phik}
  \Phi_{\text{uc}}(\ell^{(k)} \le \ell, N) = 
    \exp(-m) \sum_{i=0}^{k-1} \frac{m^{i}}{i!},
\end{equation}
where
\begin{equation}
  m = -\log(\Phi_{\text{uc}}(\ell^{(1)} \le \ell | N)).
\end{equation}

\section{DETERMINATION OF CONNECTIVITY CONSTANTS}\label{Sec:Mu}

Let $C_{r}(N)$ be the number of $N$-step random walks in which no loop of size
{\em greater than\/} $r$ is formed. The case $r = 0$ corresponds to
self-avoiding walks. As the total number of random walks of $N$-steps is $4^{N}$
on square lattice, we have
\begin{equation}\label{E:PhiC}
  \Phi(\ell^{(1)} \le r | N) = \frac{C_{r}(N)}{4^{N}}.
\end{equation}
For large $N$ it is expected that \cite{con96}
\begin{equation}\label{E:CMu}
  C_{r}(N) \sim \mu_{r}^{N}.
\end{equation}
For large $N$, $\mu_{r}$ tends to a constant independent of $N$, which may be
called the {\em ``$r$-th connectivity constant''\/}. From Jensen and Guttmann
\cite{jen99} the value of $\mu_{0}$ is known very precisely and we have
estimated $\mu_{2}$ and $\mu_{4}$ using series expansion and exact enumeration
(details follow).

Now consider the case $k = 2$. In this case, the walk cannot form loops, except
that it can go back along the path it has taken. The connectivity constant
$\mu_{0}$ can be interpreted as the average number of forward directions
available for the next step in an $N$-step self-avoiding walk for large $N$. As
return along the direction of last step is now allowed, in the first
approximation, we should have
\begin{equation}\label{E:mu2mu0ge}
  \mu_{2} \approx \mu_{0} + 1
\end{equation}
We also have the trivial inequality $\mu_{r} < \mu_{r+2}$ for all $r$. As $r$
tends to infinity, $\mu_{r}$ tends to $4$.

We determined the numbers $C_{r}(N)$ for $N \le 20$ and for all $r$ by exact
enumeration. The enumeration results for $r = 2$ and $4$ are tabulated in
Table~\ref{T:Walks}. We analyzed this data by fitting it to the extrapolation
form 
\begin{eqnarray}
\nonumber
  C_{r}(N)
    & = & K_{0} \mu_{r}^{N}(N) N^{\gamma-1}
\\
\label{E:CnMu}
    &   & \times
          \left[
	    1 + \frac{K_{1}}{N}
	    + \frac{(-1)^{N}}{N^{\gamma+1/2}}
	    \left\{
	      K_{2} + \frac{K_{3}}{N}
	    \right\}
          \right],
\end{eqnarray}
where the critical exponent $\gamma$ is expected to be independent of $r$ and
takes the self-avoiding walk value of $43/32$ in two dimensions \cite{con96} and
$K_{i}$ are constants which depend on $r$. This form is similar to that used by
Conway and Guttmann \cite{con96} for analyzing $51$-term series of self-avoiding
walks. We have reduced the number of parameters in Eq.~(\ref{E:CnMu}) because
our series is shorter. Our estimates of $\mu_{2}$ and $\mu_{4}$, by fitting the
form given by Eq.~(\ref{E:CnMu}) term-by-term to the $20$-term series tabulated
in Table~\ref{T:Walks}, are $3.7083(2)$ and $3.8818(4)$, respectively. These
values are not very sensitive to variation in the fitting values of the
parameters $K_{i}$.

It is interesting to compare the numerical values of $\mu_{0}$, $\mu_{2}$, and
$\mu_{4}$ with the estimates obtained using the uncorrelated approximation. From
Eqs.~(\ref{E:PhiC}) and (\ref{E:CMu}) we see that $\Phi(\ell^{(1)} \le \ell |
N)$ varies as $(\mu_{\ell}/4)^{N}$ for large $N$. Thus the approximation
Eq.~(\ref{E:PhiExp}) gives $\mu_{k}/4 \approx 1 - F(k)$. Using the values of
$\mu_{k}$ determined above, this would imply that $F(0)$, $F(2)$, and $F(4)$
have the values $0.3404$, $0.0729$, and $0.0295$, respectively. The values of
these quantities obtained from simulations are $0.3125$, $0.0625$, and $0.0257$,
respectively. We see that the approximation fares rather well in relating the
properties of the self-avoiding and loop-erased walks, which have quite
different large-scale properties.

\section{COMPUTER SIMULATION RESULTS}\label{Sec:Sim}

We generated two-dimensional loop-erased random walks using the algorithm
outlined in \cite{hadd01}. For each walk we collected statistics about the
perimeter and the area of the erased-loop at each step. The statistics were
collected for $N$-step walks with $N = 2^{r}$, $r = 15$, $\cdots$, $20$. We
averaged over $4.7 \times 10^{5}$ different realizations of the random walk. We
were able to simulate the entire ensemble in about $93$ hours on a Pentium-III
700 MHz machine using about $2.6$ Mb RAM.

\subsection{Largest loop perimeter}\label{Sec:Per}

During the simulations we collected statistics for $\tilde{F}(\ell | N)$, the
average number of loops of perimeter $\ell$ formed from a random walk of $N$
steps. For each walk we also determined the perimeter and area of the five
largest loops formed. this is used to obtain the measured cumulative
distribution $\Phi_{\text{o}}(\ell^{(k)} \le \ell | N)$, of size of loops of
rank $k$, with $k = 1$ to $5$. The subscript ``o'' here refers to ``observed''.
To reduce noise, nearby $\ell$ values were binned together. We used $30$ bins
per decade of data. 

\begin{figure}
\centerline{\epsfbox{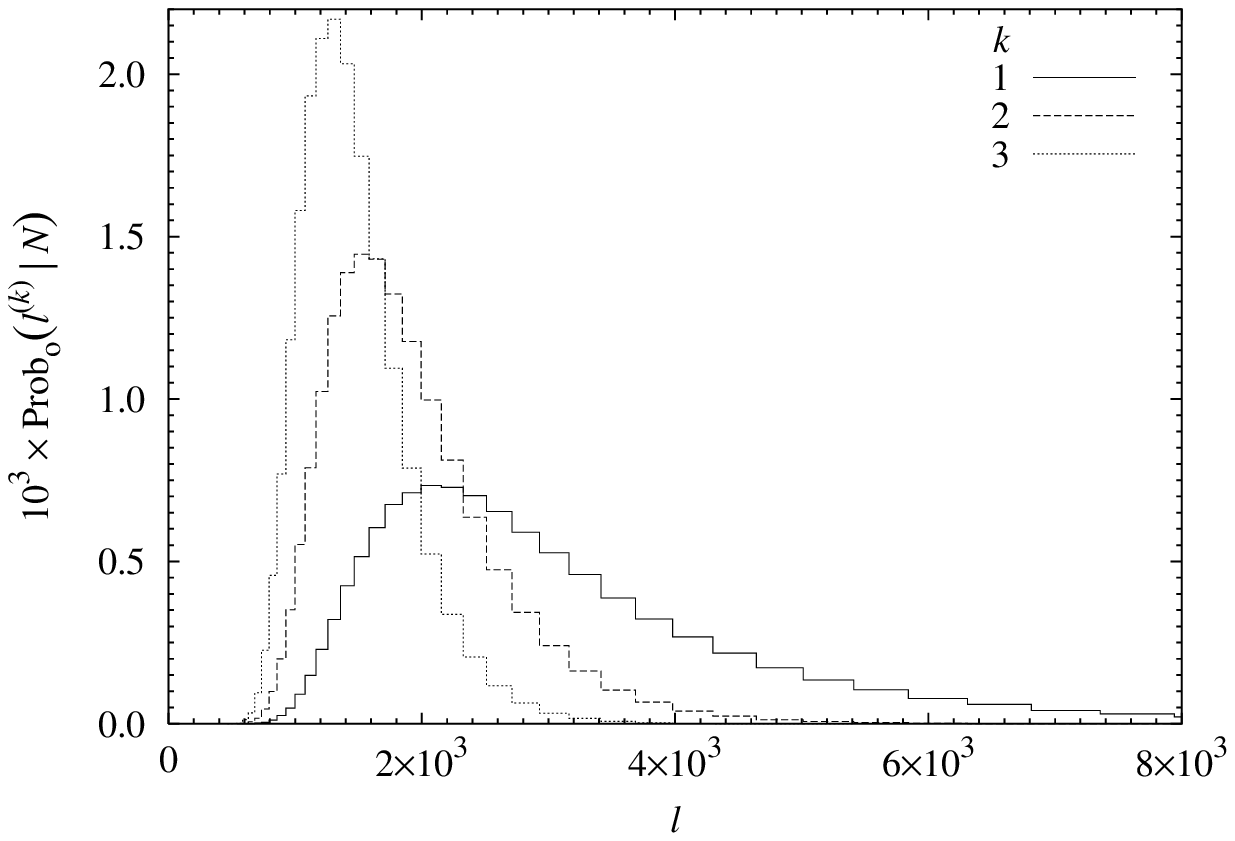}}
\vskip1ex

\caption{The observed probability distributions for perimeter of the $k$th
largest erased-loop, $k = 1$, $2$, and $3$, for two-dimensional LERW for $N =
2^{20}$.}

\label{F:llpdist15}
\end{figure}

\begin{figure}
\centerline{\epsfbox{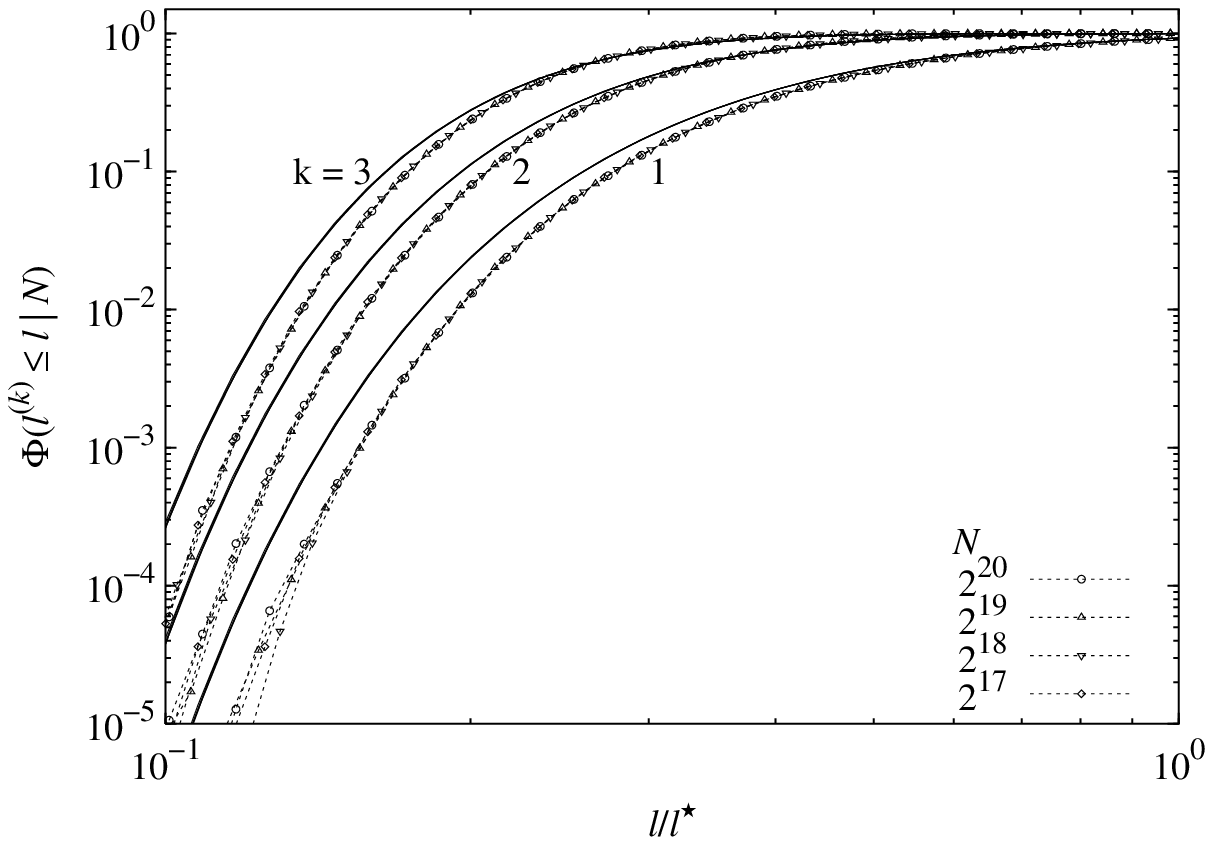}}
\vskip1ex

\caption{The cumulative probability distribution for perimeter of the $k$th
largest erased-loop, $k = 1$, $2$, and $3$, for different values of $N$ for
two-dimensional LERW. Solid lines give the prediction of the uncorrelated theory
and dashed lines with symbols give the numerically observed distributions. For
$\ell/\ell^{\star} > 1$ the curves match well with $\Phi(\ell^{(k)} \le \ell |
N)$ approaching unity very fast. Note the excellent collapse of the lines of the
same type for all values of $N$ and $k$ and also the systematic deviation (over
prediction) of the uncorrelated theory from the numerically observed
distribution.}

\label{F:llpcdist15}
\end{figure}

In Fig.~\ref{F:llpdist15} we have shown the plot for $\text{Prob}_{\text{o}}
(\ell^{(k)} | N)$ versus $\ell$ the observed probability distributions for $k =
1$, $2$, and $3$ for $N=2^{20}$. In Fig.~\ref{F:llpcdist15} we have plotted
$\Phi(\ell^{(k)} \le \ell | N)$ versus $\ell/\ell^{\star}$ for various values of
$N$ as found in the simulations, and compared it to the theoretical curve given
by Eq.~(\ref{E:phik}) ignoring correlations between loops. An excellent collapse
is seen among curves for all the values of $N$ when plotted against the scaling
variable $x = \ell/\ell^{\star}$. From these figures it is clearly seen that for
$x > 1$ the prediction of the uncorrelated theory is quite good and indeed
asymptotically exact. However, considerable departure is seen for smaller values
of $x$, for $x \ll 1$.

We see that the prediction of the cumulative distribution function by the
uncorrelated theory is consistently higher compared to the observed distribution
throughout the range of variation of the scaling variable $x$. This shows the
expected anti-correlation between occurrences of large loops.

\begin{figure}
\centerline{\epsfbox{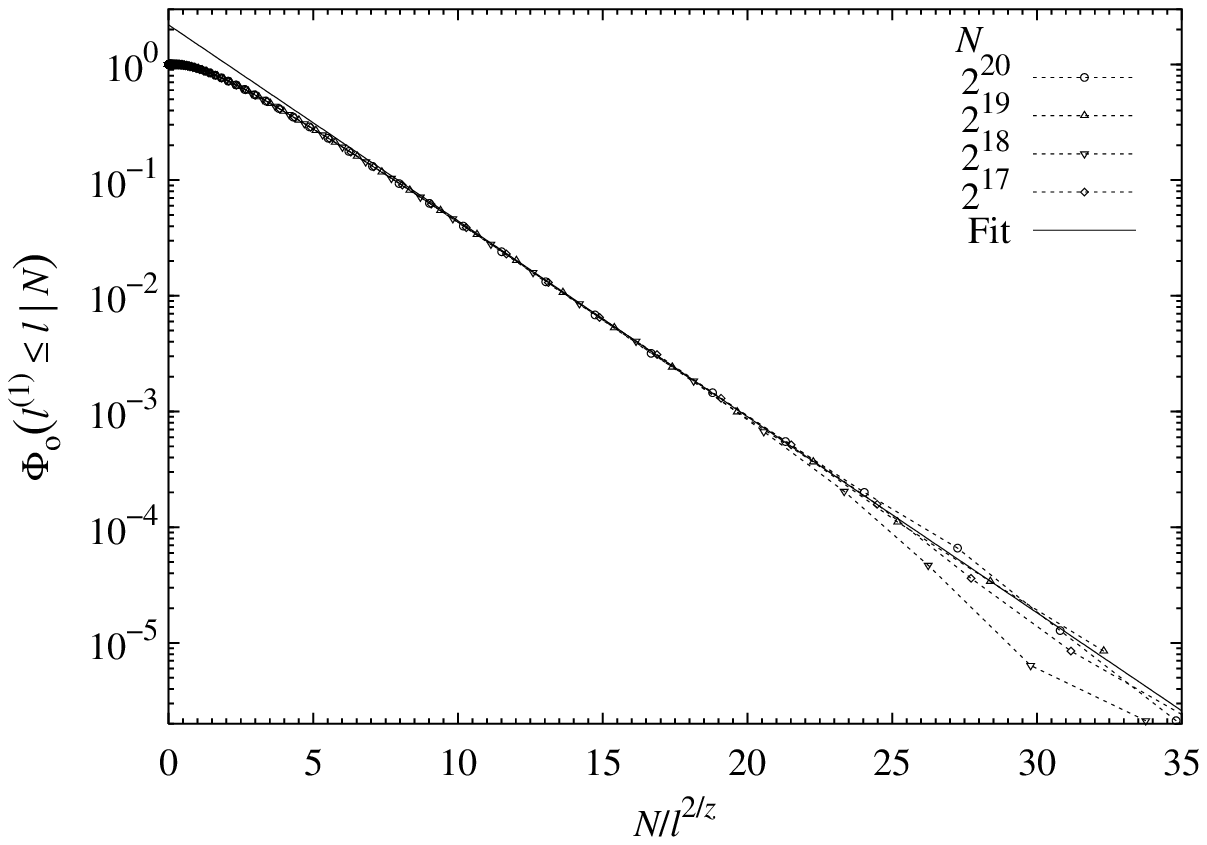}}
\vskip1ex

\caption{Variation of the cumulative probability distribution for perimeter of
the largest erased-loop for small $\ell$ for different values of $N$ for
two-dimensional LERW. Solid line gives curve-fit corresponding to
Eq.~(\ref{E:PhiOSmallL}) and dashed lines with symbols give the numerically
observed distributions.}

\label{F:llpcsl}
\end{figure}

\begin{figure}
\centerline{\epsfbox{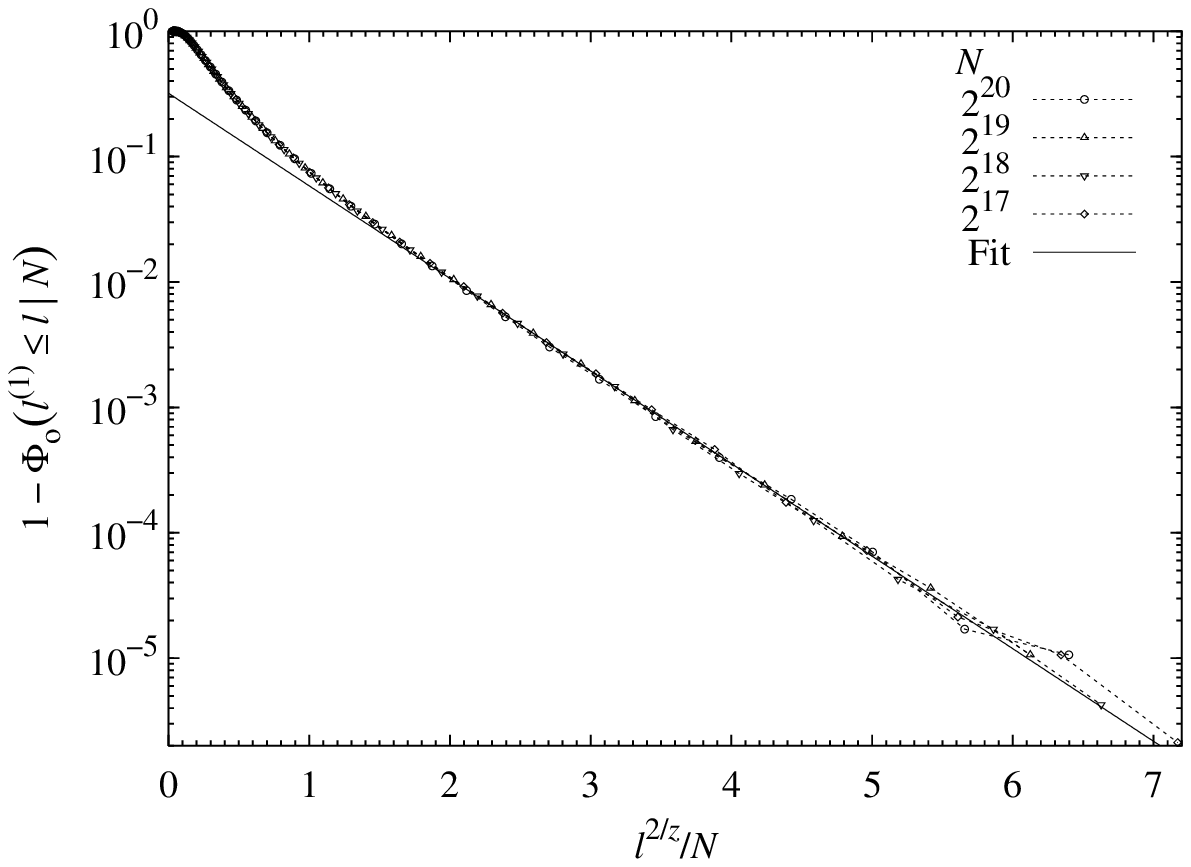}}
\vskip1ex

\caption{Variation of the cumulative probability distribution for perimeter of
the largest erased-loop for large $\ell$ for different values of $N$ for
two-dimensional LERW. Solid line gives curve-fit corresponding to
Eq.~(\ref{E:PhiOLargeL}) and dashed lines with symbols give the numerically
observed distributions.}

\label{F:llpcll}
\end{figure}

For small values of the scaling parameter $x$, the observed cumulative
distribution function seems to behave like
\begin{equation}\label{E:PhiOSmallL}
  \Phi_{\text{o}}(x^{(1)} \le x) \sim a\exp(-b x^{-2/z})
\end{equation}
with $a = 2.2 \pm 0.3$ and $b = 0.39 \pm 0.02$. The fit is shown in
Fig.~\ref{F:llpcsl}. For large $x$, $1-\Phi_{\text{o}}(x^{(1)} \le x)$ is very
nearly $N \tilde{F}(\ell | N)$ which varies as
\begin{equation}\label{E:PhiOLargeL}
  1-\Phi_{\text{o}}(x^{(1)} \le x) \sim a\exp(-b x^{2/z})
\end{equation}
with the numerical value of the parameters obtained by curve-fitting being $a =
0.32 \pm 0.03$ and $b = 1.7 \pm 0.1$, same as that obtained by analysis of the
all-loops data. This fit is shown in Fig.~\ref{F:llpcll}.

\subsection{Largest loop area}\label{Sec:Area}

During simulations we collected statistics for the area of the erased-loops
also. This was sampled exactly as the perimeter data in the previous subsection.

\begin{figure}
\centerline{\epsfbox{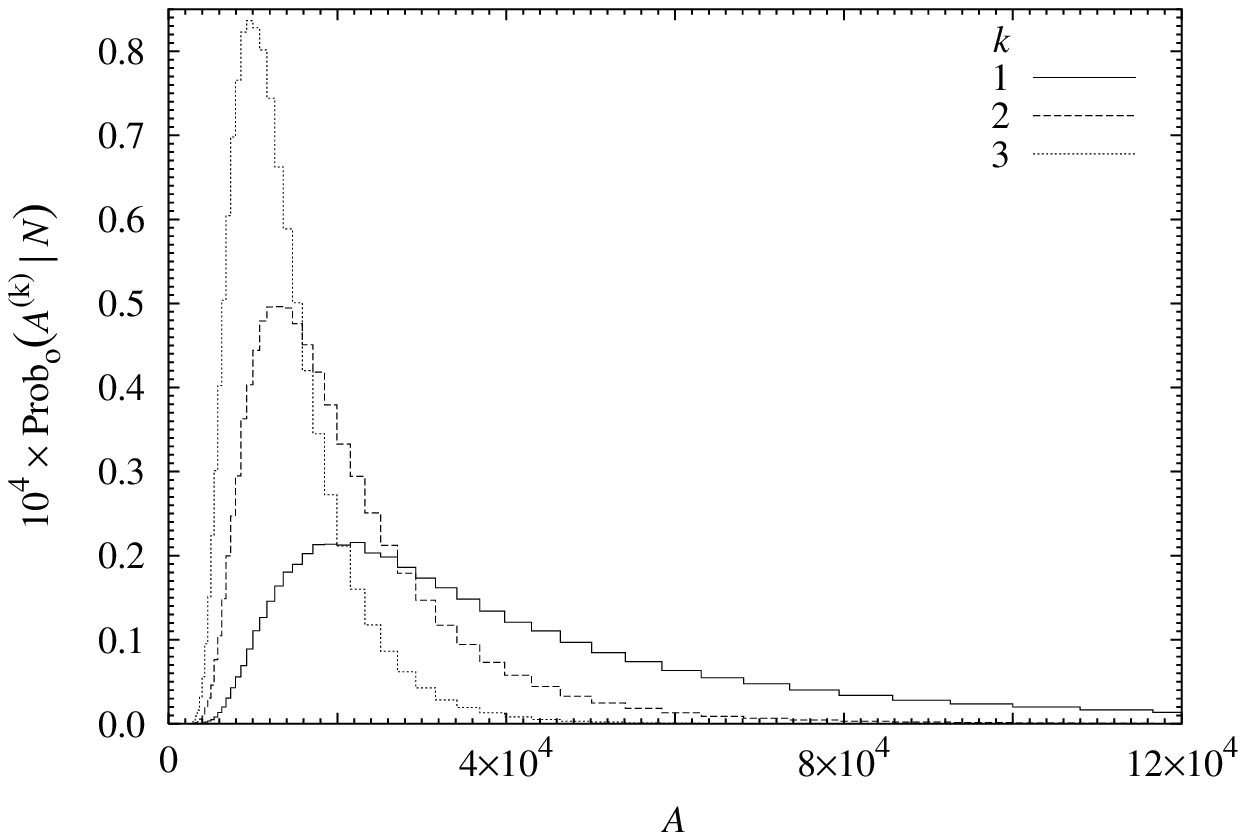}}
\vskip1ex

\caption{The observed probability distributions for area of the $k$th largest
erased-loop, $k = 1$, $2$, and $3$, for two-dimensional LERW for $N = 2^{20}$.}

\label{F:lladist15}
\end{figure}

\begin{figure}
\centerline{\epsfbox{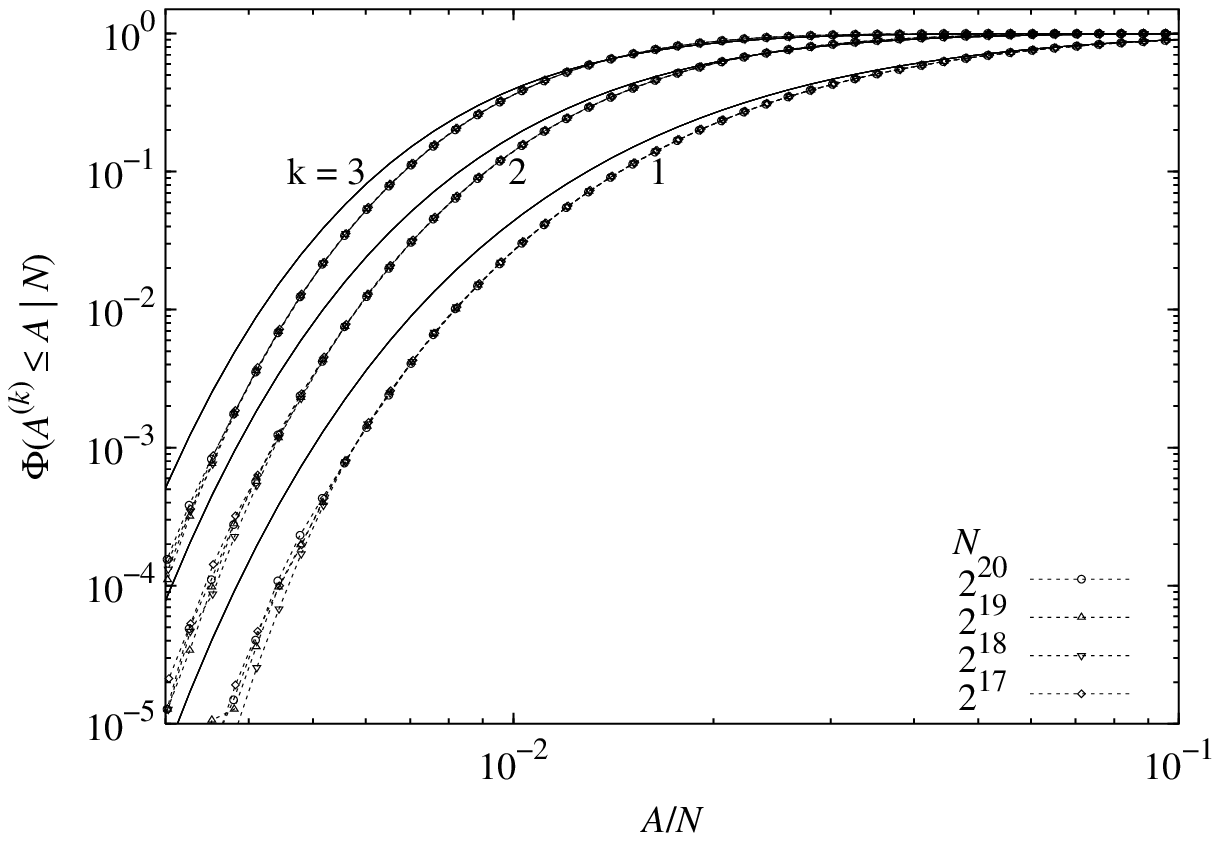}}
\vskip1ex

\caption{The cumulative probability distribution for area of the $k$th largest
erased-loop, $k = 1$, $2$, and $3$, for different values of $N$ for
two-dimensional LERW. Solid lines give the prediction of the uncorrelated theory
and dashed lines with symbols give the numerically observed distributions. For
$A/N > 0.1$ the curves match well with $\Phi(A^{(k)} \le A | N)$ approaching
unity very fast. Note the excellent collapse of the lines of the same type for
all values of $N$ and $k$ and also the systematic deviation (over prediction) of
the uncorrelated theory from the numerically observed distribution.}

\label{F:llacdist15}
\end{figure}

In Figs.~\ref{F:lladist15} and \ref{F:llacdist15} we have shown the plots for
$\text{Prob}_{\text{o}}(A^{(k)} | N)$ versus $A$ for $N=2^{20}$ and
$\Phi(A^{(k)} \le A | N)$ versus $A/N$ for various values of $N$, for $k = 1$ to
$3$. The format of presentation is as in the previous subsection. An excellent
collapse is seen among the curves for various values of $N$ when plotted against
the scaling variable $y = A/N$. 

The departure between the observed behavior and prediction of the uncorrelated
theory is also similar to that seen for the perimeter data in the previous
subsection. It is clearly seen from these figures that for $y > 0.1$ the
prediction of the uncorrelated theory is quite good and seems to be
asymptotically exact for large $y$. For $y < 0.1$ considerable departure is seen
between observed behavior and uncorrelated prediction. As in the perimeter data,
there is a systematic over-prediction by the uncorrelated theory.

For small values of the scaling parameter $y$, the observed cumulative
distribution function $\Phi_{\text{o}}(y^{(1)} \le y)$ seems to behave like
$\exp(-a/y)$ with $a = 0.049 \pm 0.002 $. For large $y$,
$1-\Phi_{\text{o}}(y^{(1)} \le y)$ varies as $\exp(-by)$ with $b = 14 \pm 1$.

\subsection{Variation of loop sizes with rank}\label{Sec:kth}

It is clearly seen in Fig.~\ref{F:llpdist15} that the probability distribution
of $\ell^{(k)}$ becomes sharper as $k$ increases. In fact, if $k$ is of order
$N$ (say $k = N/1000$), it is easy to see that the distribution tends to a delta
function for large $N$. A more careful argument shows that if $k \gg
N^{z/(z+1)}$, then the distribution would tend to a delta function. We note that
$\ell^{(k)}$ varies as $(N/k)^{z/2}$ and the average number of erased loops with
this perimeter varies as $N/(\ell^{(k)})^{1+2/z}$. For the distribution to have
sharp peak at $\ell^{(k)}$, this number should be much greater than fluctuations
in the expected number of loops with perimeter greater than $\ell^{(k)}$. The
latter varies as $k^{1/2}$. Simple algebra then gives the required result. 

A similar argument for the area distributions shows that the positions of the
peak for the $k$-th rank varies roughly as $N/k$ and their width varies as
$N/k^{3/2}$. Furthermore, when $k \gg N^{2/3}$ the width of the distribution
becomes exponentially small in $N$.

\subsection{Affect of correlations on the probability distribution functions for
$k$th largest erased-loop size}
\label{Sec:Affect}

\begin{figure}
\centerline{\epsfbox{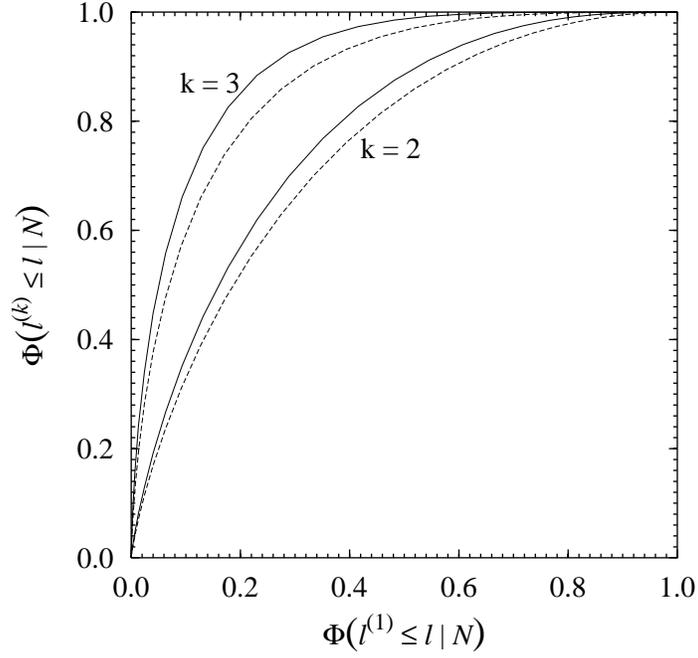}}
\vskip1ex

\caption{Variation of the cumulative probability distribution for perimeter of
the $k$th largest erased-loop, $k = 2$ and $3$, with that of the largest
erased-loop for two-dimensional LERW. Dashed lines give the prediction by
uncorrelated theory and solid lines give the behavior of the observed data. Here
the curves are shown only for $N=2^{20}$. Curves for other values of $N =
2^{r}$, $r = 17, 18, 19$, collapse indistinguishably with these curves.} 

\label{F:llpc12}
\end{figure}

In Fig.~\ref{F:llpc12}, we have plotted $\Phi(\ell^{(2)} \le \ell | N)$ and
$\Phi(\ell^{(3)} \le \ell | N)$ versus $\Phi(\ell^{(1)} \le \ell | N)$ for $N =
2^{20}$ from the observed distributions. This is compared with what would be
expected on the basis of uncorrelated approximation. Similar plots using area
(instead of perimeter) data show similar trends, and are omitted here. From this
figure, it is clearly seen that the predicted and the observed distributions are
quite close. The actual curve always lies above the value calculated by
neglecting anti-correlations present.

\section{MODELING CORRELATIONS}\label{Sec:Corr}

Consider the time series $\{n_{i}\}$ with $i = 1$, $2$, \ldots, generated in a
LERW simulation, where $n_{i}$ is number of steps in the LERW at time step $i$.
This process can be modeled by a stochastic motion of point on a one dimensional
lattice. As $n_{i}$ is always positive, the motion occurs in the half space $x
\ge 0$. In a single time step, this point can move one step to the right (if no
loop erasure occurs in the corresponding random walk), or several spaces to the
left. Now suppose that the random walk is not accessible to observation, and
only the time series $\{n_{i}\}$ is observed. While the original LERW, treated
as a stochastic process is a Markov process, the projected process is clearly
{\em not\/} Markovian. However, it may be approximated as a Markov process. 

\subsection{One-dimensional Levy walk model}

The transition probabilities for this Markov process are easily defined. We
think of $n_{i}$ as the position of a random walker at time $i$ on a one
dimensional lattice. The walk begins at $t = 0$ with the walker positioned at $x
= 0$. At each subsequent time step, the walker takes one step to the right and
then draws a non-negative integer random number $\ell$ with the probability
$\text{Prob}(\ell)$, $\ell=0$, $1$, $2$, \ldots. We will assume that for large
$\ell$, $\text{Prob}(\ell)$ decreases as $\ell^{-\tau}$ with $\tau > 1$. If
$\ell$ is less than or equal to the current position $x$ of the walker, then the
walker takes $\ell$ steps to the left; otherwise it stays put. This completes
one step. Clearly, we have
\begin{equation}
\label{E:plsum}
  \sum_{\ell = 0}^{\infty} \text{Prob}(\ell) = 1.
\end{equation}
To ensure that there is no overall drift in the model, we also assume that
\begin{equation}\label{E:firstmo}
  \sum_{\ell = 0}^{\infty} \ell \text{Prob}(\ell) = 1.
\end{equation}

Note that the $\ell$ here corresponds  to the erased-loop size in LERWs. In
general, one can expect to improve comparison with the original LERW model by
making the probability of backward $\ell$ steps when the walker is at $n$ equal
to the conditional probability in the LERW problem that the next step leads to
erasure of a loop of length $\ell$ when the current length of walk is $n$. This
is expected to be of the form
\begin{equation}
  \text{Prob}(\ell | n) = \text{Prob}(\ell | \infty) f_{\text{cutoff}}(\ell/n),
\end{equation}
where $f_{\text{cutoff}}$ is a cutoff function which is strictly zero if its
argument is greater than $1$. We make the simple choice that $f_{\text{cutoff}}$
is $1$ if the argument is less than $1$.

For our simulations, we made a particular choice of $\text{Prob}(\ell)$. We
assumed that it is given by
\begin{equation}\label{E:Pell}
  \text{Prob}(\ell) =
    \left\{
    \begin{array}{ll}
    \displaystyle
      \frac{1}{\ell} \left[ \frac{1}{\ell^{\alpha}} - \frac{1}{(\ell+1)^{\alpha}} \right]
      ,
	& \quad \text{for~} 1 \le \ell \le \infty
\\\\
    \displaystyle
      1 - \sum_{k = 1}^{\infty} \text{Prob}(k)
      ,
	& \quad \text{for~} \ell = 0.
    \end{array}
    \right.
\end{equation}
Note that with this choice, the no-drift condition given by
Eq.~(\ref{E:firstmo}) is automatically guaranteed for any choice of $\alpha$.
Furthermore, one can generate this distribution numerically by using only two
calls to the random number generator. We take a random number $u$ with uniform
distribution between $[0,1]$, define $m = \lfloor u^{-1/\alpha}\rfloor$, and
then put $\ell = m$ with probability $1/m$ and $\ell = 0$ with probability
$1-1/m$. In our simulations, we used $\alpha = 0.6$, which corresponds to the
value $\tau = 2.6$ of the exponent of the two-dimensional LERWs.

The Master equation for the above process describing the evolution of the
probability $P(x,t)$ of the walker being at position $x$ at time $t$ is written
as
\begin{equation}\label{E:ME}
  P(x,t+1) = \sum_{\ell = 0}^{\infty} \text{Prob}(\ell) P(x - 1 + \ell,t).
\end{equation}

For large times $t$, the width of the probability distribution $P(x,t)$
increases to infinity. It is easy to see that the width must increase as
$t^{1/(1 + \alpha)}$. We note that if the particle it at $x$, its expected
displacement in the next time-step is positive, as jumps with displacement
greater than $x$ to the left are disallowed. The contribution of such terms to
Eq.~(\ref{E:plsum}) varies as $x^{2-\tau}$. This equation may schematically be
written in the form
\begin{equation}\label{E:Diff}
  \frac{\partial P}{\partial t}
    \sim \frac{\partial}{\partial x}(P x^{2-\tau}) + {\cal D}P,
\end{equation}
where ${\cal D}$ denotes diffusion operator which, presumably, involves
fractional derivatives. The resulting equation for the scaling function is
nonlocal, and its analytical solution seems difficult. Simple dimensional
analysis shows that $t$ scales as $x^{\tau-1}$. Hence the width of this
distribution should scale as $t^{1/(\tau-1)}$. Furthermore, for large $t$,
$P(x,t)$ tends to the scaling form
\begin{equation}\label{E:Pxt}
  P(x,t) \simeq
    \frac{1}{t^{1/(\tau -1)}} p\left(\frac{x}{t^{1/(\tau-1)}}\right).
\end{equation}

\subsection{Results from the Levy walk model}

We numerically integrated the Master equation Eq.~(\ref{E:ME}) in $x \ge 0$ half
space using the probability distribution for erased-loop sizes given by
Eq.~(\ref{E:Pell}) and computed $P(x,t)$. The integration for walks having up to
$N = 2^{17}$ steps required about $80$ hours of CPU time on a Pentium II 350 MHz
machine using about $7$ Mb RAM. We also simulated the Levy walk process for time
steps up to $N = 2^{20}$ for obtaining the statistics on erased-loop sizes and
the $k$th largest erased-loop size. The quantities were sampled along the same
lines as for the LERWs discussed in Sec.~\ref{Sec:Sim}. To reduce noise in the
statistics, we averaged over a large ensemble consisting of $2 \times 10^{5}$
different runs. The simulation of the entire ensemble required about $141$ hrs
of CPU time on a Pentium II 350 MHz machine using about $1.5$ Mb RAM.

\begin{figure}
\centerline{\epsfbox{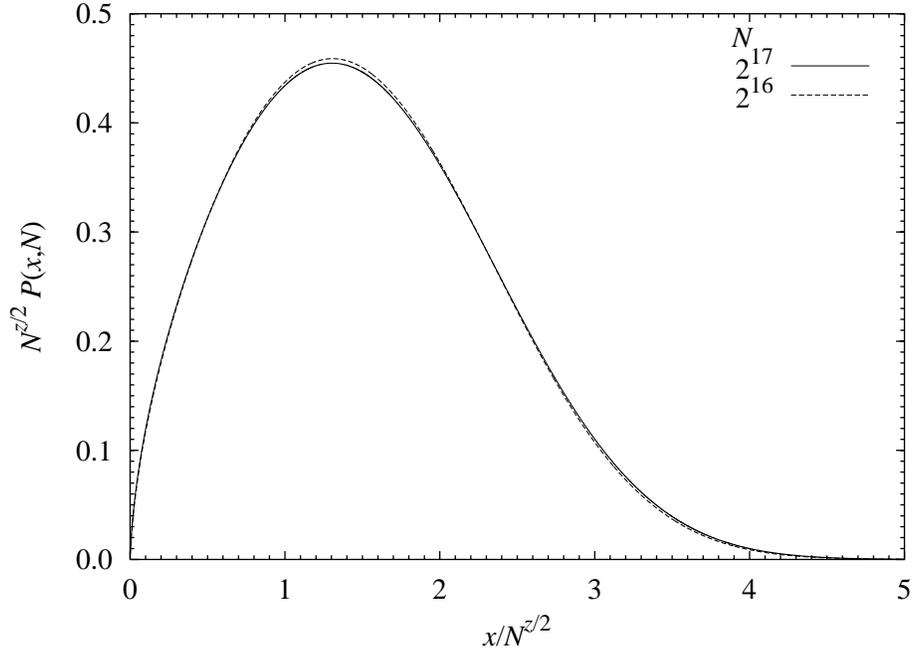}}
\vskip1ex

\caption{Scaling plots from numerical integration of the Master equation
Eq.~(\ref{E:ME}) for probability of finding the Levy walker at position $x$ at
time step $N$ versus $x/N^{z/2}$, $z = 5/4$, for $N = 2^{16}$, and $2^{17}$.
Good scaling and consequently good collapse of curves is seen.}

\label{F:pxdist}
\end{figure}

Scaling plots for the computed probability of finding the Levy walker at
location $x$ at time step $N$, $P(x,N)$, are shown in Fig.~\ref{F:pxdist}. In
this figures we have plotted $N^{z/2} P(x,N)$ versus $x/N^{z/2}$, for $z=5/4$.
The figure clearly shows that the observed behavior agrees well with the
conjectured scaling form given by Eq.~(\ref{E:Pxt}).

We also analyzed the distribution of $k$th largest loop sizes in  simulation of
this Levy walk model, and compared them with the corresponding distributions for
the two-dimensional LERW model. We found that the deviations from the
predictions of the uncorrelated theory are much smaller in the case of the Levy
walk model than in the original LERW. The plots are very similar to the
Figs.~\ref{F:llpdist15}, \ref{F:llpcdist15}, and \ref{F:llpc12}, and are being
omitted here.

\begin{figure}
\centerline{\epsfbox{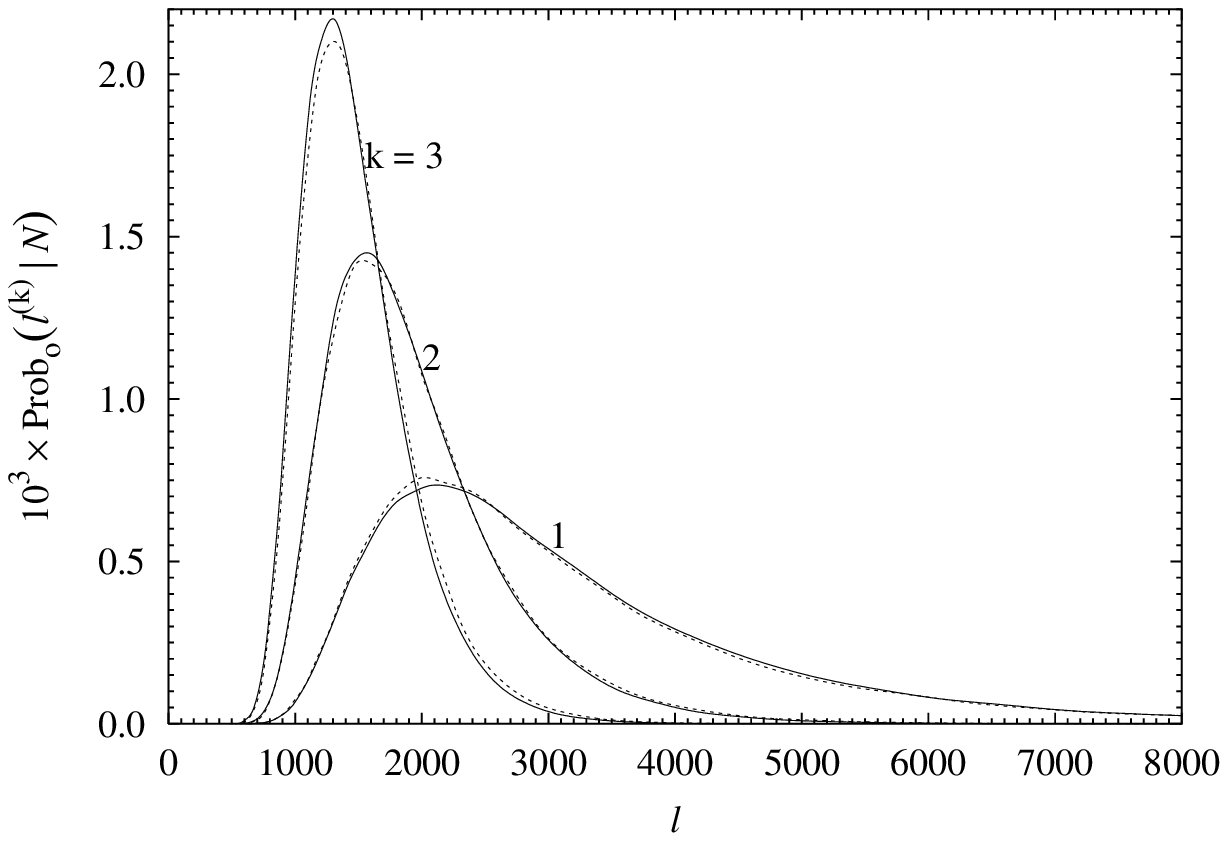}}
\vskip1ex

\caption{Observed probability distributions for size (perimeter for LERW) of the
$k$th largest erased-loop for two-dimensional LERW (solid lines) and the Levy
walk model (dashed lines) for $N = 2^{20}$. The extremal distributions for the
Levy walk model have been rescaled by multiplying (dividing) the abscissa
(ordinate) by a factor of $1.04$. This rescaling makes the mean points of the
distributions obtained from the Levy walk model coincide with those of the
LERW.}

\label{F:lw1le2distR}
\end{figure}

In Fig.~\ref{F:lw1le2distR}, we have compared the probability distributions for
the $k$th largest erased-loop sizes from the Levy walk model with those from
LERW. The figure clearly shows that the probability distributions obtained from
the Levy walk model match very well with those from the LERW.

A better quantitative estimate can be obtained by comparing the ratio $R_{k}$,
defined as
\begin{equation}
  R_{k} = \langle\ell^{(k)}\rangle / \langle\ell^{(1)}\rangle, 
\end{equation}
where $\langle \rangle$ denotes expectation value.

The value of $R_{k}$ as found in the simulations of the LERW was found to be
$0.605$, $0.463$, $0.386$, and $0.335$ for $k = 2$ to $5$, respectively. The
corresponding values in the simulation of the Levy walk model were $0.614$,
$0.474$, $0.397$, and $0.347$, respectively. The corresponding values from the
uncorrelated approximation would be ${k}^{-z/2}$, i.e., $0.648, 0.503, 0.420$,
and $0.365$, respectively. It is clear that the Levy walk model gives a much
better estimate of these ratios than the uncorrelated approximation.

\section{CONCLUDING REMARKS}\label{Sec:Conc}

Our analysis above shows that the probability distribution of the largest
erased-loops in LERWs is fairly well described by the simple approximation
ignoring correlations between the sizes of different loops. However, the average
values of ratios of $\ell^{(k)}$ are not well described in this approximation. A
simple model which takes care of a large part of these correlations is the Levy
walk model introduced in this paper. In this model, one keeps information about
the {\em length\/} of the LERW, but throws out all information about its shape.
We have seen that this model reproduces the extremal statistics of the LERWs
quite well.

Secondly, we have exactly enumerated $C_{r}(N)$ the number of $N$ step LERWs in
which loops of size {\em less than or equal\/} to $r$ are erased. Using these we
have determined $\mu_{r}$ the $r$th connectivity constant. The determination of
$\mu_{0}$ for various lattices has been a long-standing problem in lattice
statistics. Higher $r$-values present interesting geometrical questions, and may
be helpful in understanding the crossover from random walk to self-avoiding
walk. 

%\section*{ACKNOWLEDGEMENTS}

%We would like to thank M. Barma for his critical reading of this paper.

\newpage

\begin{table}
\caption{Number of $N$-step loop-erased random walks $C_{\ell}(N)$ in which the
largest loop of perimeter $\ell$ less than or equal to $2$ and $4$ are erased
for $N = 1$, $\cdots$, $20$.}
\vskip1ex
\label{T:Walks}

\begin{tabular}{rrr}
$N$ & $C_{2}(N)$   & $C_{4}(N)$   \\\hline
 1     &            4 &            4 \\
 2     &           16 &           16 \\
 3     &           64 &           64 \\
 4     &          248 &          256 \\
 5     &          976 &         1024 \\
 6     &         3736 &         4072 \\
 7     &        14536 &        16248 \\
 8     &        55280 &        64352 \\
 9     &       213336 &       256120 \\
10     &       808016 &      1011504 \\
11     &      3099456 &      4016496 \\
12     &     11706568 &     15828968 \\
13     &     44696992 &     62727520 \\
14     &    168475176 &    246805224 \\
15     &    640913784 &    976340664 \\
16     &   2411998168 &   3836482296 \\
17     &   9148925856 &  15153764480 \\
18     &  34387933200 &  59482843856 \\
19     & 130125970320 & 234640138528 \\
20     & 488603502672 & 920216177360 \\
\end{tabular}
\end{table}

\onecolumn
\widetext

\begin{references}
\bibitem{gumnote} To be more precise, the three extreme value distributions were
discovered by Fr\'echet, and Fisher and Tippett, and discussed more completely
by Gnedenko \cite{gned43}.
\end{references}
\end{document}